\documentclass[12pt, letterpaper]{article}
\usepackage[utf8]{inputenc}
\usepackage{amsmath}
\usepackage{graphicx}
\usepackage{authblk}
\graphicspath{ {./images/} }
\usepackage[utf8]{inputenc}
\usepackage{amssymb}
\usepackage{multicol}
\usepackage{dcolumn}
\usepackage{changepage}
\begin{document}



\title{Lorentz-violating pseudovectors in effective field theories for quantum gravity}
\author{Hollis Williams* }

\affil[$1$]{Department of Physics, University of Warwick \\ Coventry CV4 7AL, United Kingdom

*holliswilliams@hotmail.co.uk}

\maketitle

\begin{abstract}

Effective field theories which describe the coupling between gravity and matter fields have recently been extended to include terms with operators of non-minimal mass dimension.  These terms preserve the usual gauge symmetries but may violate local Lorentz and diffeomorphism invariance.  The number of possible terms in the field theory explodes once one allows for non-minimal operators, with no criterion to choose between them.  We suggest as such a criterion to focus on terms which violate Lorentz invariance via a (pseudo)vector background field, leaving a number of possible terms in the Higgs, gauge and gravitational sectors.  Further study of these terms is motivated by the proposed correspondence between the general effective theory for Lorentz violation and emergent Lorentz symmetry in condensed-matter systems, which is mostly unexplored for higher mass dimension operators and couplings to gauge fields and gravity.  We suggest bounds in the Higgs sector and we show that some of the coefficients in the gauge sector vanish at one loop, whereas others have bounds which are comparable with those suggested by Kostelecký and Li for coefficients in Lorentz-violating QCD and QED coupled to quarks.  We also find new bounds in the gravitational sector by considering Robertson-Walker cosmology.  Finally, we discuss the special case where only diffeomorphism invariance is spontaneously broken and explain why it does not allow for non-trivial Nambu-Goldstone modes.    


\end{abstract}

\section{Introduction}
\label{}
\noindent
The search for a compelling theory which unifies classical gravity with quantum mechanics is a major open problem in theoretical physics.  The nature of this theory is currently mysterious and may involve deviations from non-Riemannian geometry.  The underlying theory could even be non-geometric and so completely distinct from the gravitational theories which we use now [1].  Given this uncertainty, it is clearly of interest to study model-independent signals from the Planck scale which trickle down to the low energies which we observe in the world around us.  It should be possible (at least in principle) to determine certain properties of the underlying theory by performing experiments at low energies.  One way that this can be done is to use the machinery of effective field theory [2].  In this approach, one starts with the Lagrangian of general relativity coupled to the Standard Model and adds correction terms which are products of operators built from known fields multiplied by coefficients associated with the underlying theory.  Background fields in the underlying theory are reflected at the level of the EFT as coupling coefficients.  Coupling coefficients in EFTs are usually assumed to be constant scalars, but if the unifying theory has more complicated backgrounds, these may manifest themselves in the EFT as vector or tensor functions which depend on spacetime position.  Tensor background fields which are included as couplings in the EFT can transform anomalously under spacetime transformations, so that the EFT may have apparent or genuine violations of both local Lorentz invariance and diffeomorphism invariance.  

The general framework to construct an EFT which describes GR and the SM along with possible violations of this kind was found by Colladay and Kostelecký (known as the Standard Model Extension)  [3].  In this framework, terms in the EFT are arranged into a hierarchy depending on the mass dimension $d$.  Minimal terms are considered to be those whose background fields have mass dimension $d \leq 4$.  We emphasise that we are referring to the mass dimensions of the backgrounds (the Lorentz-violating terms themselves have overall mass dimension $4$ to comply with renormalizability).  All of the possible minimal terms were found by Kostelecký  [4].  The possible terms have been studied intensively in the case of Lorentz-violating quantum electrodynamics [5].  The result for minimal terms was extended by Kostelecký  and Li to all terms with $d \leq 6$ [6].  It is important to consider non-minimal terms, since terms with larger $d$ are expected to have weaker effects at low energies, in agreement with the feebleness of quantum gravity effects at low energy.  The main problem with studying the non-minimal theory is that the number of possible terms explodes with exponential speed once we start moving to terms with coefficients of higher mass dimension, with no apparent criterion to choose between them.

In this work, we consider a special form of Lorentz-$CPT$ breaking theories of scalar, gauge and gravitational fields characterized by a single Lorentz-violating (pseudo)vector.  In terms of the theoretical reason to study these models, it should be borne in mind that the data tables of [7] include both Lorentz-invariant and Lorentz-violating terms, depending on the exact structure of the coefficients.  Operators with odd numbers of spacetime indices are CPT-odd and always break Lorentz symmetry, hence these terms are the simplest possible that must break Lorentz symmetry, and so the simplest which could be phenomenologically viable for detecting Lorentz violation [8].

Secondly, although studies of symmetry breaking are naturally motivated by searches for limits of applicability of Lorentz symmetry, they may also be used to develop effective models applied in various contexts, including condensed matter physics and phenomenology of elementary particles.  It is well known that Lorentz invariance is absent in condensed-matter systems but more specifically, a correspondence has been proposed for the SME and the description of emergent Lorentz symmetry in these systems [9, 10].  Terms from the minimal SME have been employed in condensed-matter physics, but the correspondence has not been explored for operators with mass dimension $>4$ and couplings to all gauge fields and gravity [11].  These terms may modify the Dirac operator in a large number of interesting ways which have not been considered, leading to possibility of new materials with emergent Lorentz symmetry violation.  Besides these modifications in the fermion sector, it may be possible to model other new condensed-matter systems using SME terms from the Higgs, gauge and gravitational sectors.  Several studies focus directly on CPT-odd (axial) contributions since this sector can describe the anomalous Hall effect and contribute to the chiral magnetic effect [12, 13].  As another example, CPT-odd terms in the fermion-photon sector of the minimal QED extension of the SME can be used to yield a restricted version of axion electrodynamics [14].

In summary, we will consider the Higgs, gauge and gravity couplings which are CPT-odd and which are also the simplest possible examples of such couplings which are represented by higher mass dimension operators. These terms were identified in [4] and it was argued that it would be desirable if they vanished since CPT-odd terms likely generate instabilities in the minimal theory, but the possibility of using experimental data to bound the associated coefficients was not discussed. Besides this, although it might be desirable to require CPT-odd terms to vanish when considering elementary particle physics, as stated above CPT-odd contributions may still be important for modelling condensed-matter systems [12]. In the other direction, insight into the stability or interpretation of these coefficients in the SME may be increased by studying them in the emergent Lorentz symmetry context [10].  The main aim of this article is to take a step in this direction by using theoretical predictions for various physical processes to bound these coefficients where possible and otherwise to discuss plausibility arguments on their size when this is not feasible.  In the case of the coefficient $\hat{k}_0$, a bound is not provided but we provide an argument that it vanishes at one loop.  An interesting qualitative prediction is suggested using $\hat{k}_3$, although it would require simulations to be verified.

This paper is organized as follows.  In Section 2, we study various physical processes and make predictions which enable us to bound various of the coefficients.  We note that the bounds which we achieve in the gauge sector match those which were suggested as feasible (but not proved) for coefficients in the limit of Lorentz violating QED and QCD coupled to multiple quarks [6].   The type of term we study is the simplest possible which allows for breaking of diffeomorphism invariance.  It is a general result that diffeomorphism invariance is spontaneously broken in a theory if and only if local Lorentz invariance is also spontaneously broken, so that the distinction is not important but there is a special case where diffeomorphism invariance can be spontaneously broken whilst preserving local Lorentz invariance.  In Section 3, we clarify that this would not have observable consequences in terms of the usual analysis of the associated Nambu-Goldstone modes, hence this case does not need to be considered.  We finish with conclusions in Section 4.

\section{Violation of Lorentz Invariance with a Vector Background Field}

\noindent
Whereas previous studies consider the minimal terms to be the ones where the backgrounds have mass dimension $\leq 4$, we instead include only terms where the coupling coefficients are single-index tensors.  The number of possible terms is relatively small and is restricted to the gauge, Higgs and gravity sectors.  The usual minimal gauge sector of the Standard Model Extension is reproduced but there are some additional higher dimension terms in the Higgs and gravity sectors.  In the gauge sector, we have  

\[  \mathcal{L}_{\text{gauge} }  = (\hat{k}_0 )_{\kappa} B^{\kappa} +  \big( \hat{k}_1  \big)_{\kappa}  \epsilon^{\kappa \lambda \mu \nu }  B_{\lambda} B_{\mu \nu}   + \big( \hat{k}_2  \big)_{\kappa}  \epsilon^{\kappa \lambda \mu \nu }  \text{tr} ( W_{\lambda} W_{\mu \nu} + \frac{2}{3} i g W_{\lambda} W_{\mu}  W_{\nu})    \] 

\[+ \big( \hat{k}_3  \big)_{\kappa} \epsilon^{\kappa \lambda \mu \nu } \text{tr} ( G_{\lambda} G_{\mu \nu} + \frac{2}{3} i g_3 G_{\lambda} G_{\mu}  G_{\nu} ) ,  \tag{1} \]

\noindent
where the background in the first term has mass dimension $1$ and the remaining backgrounds have mass dimension $3$.  $G_{\mu}$ is a Hermitian SU(3) adjoint matrix which describes the gauge fields for the strong interaction, $W_{\mu}$ is an SU(2) adjoint matrix describing the gauge fields for the weak interaction, and $B_{\mu}$ is the singlet hypercharge gauge field.  The corresponding field strengths for the gauge fields are denoted by $G_{\mu \nu}$, $W   _{\mu \nu}$, and $B_{\mu \nu}$.    In the Higgs sector, we have

\[   \mathcal{L}_{\text{Higgs}} = ( \hat{k}_{\phi} )^{\mu} \phi^{\dagger} i D_{\mu} \phi + \text{H.c.} + \frac{1}{2}   ( \hat{k}_{\phi D \phi} )^{\mu}  ( \phi^{\dagger} \phi ) ( \phi^{\dagger} i D_{\mu} \phi ) + \text{H.c.} , \tag{2}  \]

\noindent
where the background has mass dimension $3$ in the first two terms and mass dimension $5$ in the last two.  In the gravity sector, we have

\[   \mathcal{L}_{\text{gravity}} =  \frac{1}{2 \kappa} \bigg[ ( \hat{k}_{\Gamma} )^{\mu}  \Gamma^{\alpha}_{\mu \alpha}  +    \big( \hat{k}_{\text{CS},1}  \big)_{\kappa}   \epsilon^{\kappa \lambda \mu \nu } \eta_{ac} \eta_{bd} \big( \omega_{\lambda}^{ab} \partial_{\mu} \omega_{\nu}^{cd}    + \frac{2}{3}  \omega_{\lambda}^{ab} \omega_{\mu}^{ce} \omega_{\nu e}^d  \big) \]

\[ + \big( \hat{k}_{\text{CS},2}  \big)_{\kappa}   \epsilon^{\kappa \lambda \mu \nu }  \epsilon_{abcd} \big( \omega_{\lambda}^{ab} \partial_{\mu} \omega_{\nu}^{cd} + \frac{2}{3}  \omega_{\lambda}^{ab} \omega_{\mu}^{ce} \omega_{\nu e}^d  \big) \bigg]  ,  \tag{3} \]

\noindent
where the background in the first term has mass dimension $3$ and the Chern-Simons backgrounds have mass dimension $5$. 

We stated in the Introduction that we are considering the simplest possible case of Lorentz violation, but we should emphasize that we are not considering single-index Lorentz-violating backgrounds in full generality.  The reason is as follows. The hatted coefficients used in the equations (1)-(3) include the possibility of Lorentz- and CPT-violating terms with higher mass dimensions.  The point of the hat notation is that such a coefficient is not just a number, but may be a sum of terms with additional factors of the covariant derivative.  Inserting additional derivatives increases the mass dimension of the operators and it usually also increases the number of free Lorentz indices, but a structure of the form $\partial^{\mu} \partial_{\mu} ={\partial}^2$ can always be inserted between operators without requiring introduction of another external index.  For massless fields, the free propagator is not affected by terms which only have an additional $  {\partial}^2$, but such terms cannot generally be omitted for massive fields, nonlinear interactions, or radiative corrections (see, for example, [15]).




 

\subsection{Higgs Sector}
\noindent
We will begin by considering the Higgs sector.  Taking the Hermitian conjugate of $(1/2)  ( \phi^{\dagger} \phi ) ( \phi^{\dagger} i D_{\mu} \phi )$ only adds a total divergence after integrating by parts and commuting through with $\phi^{\dagger} \phi$.  Taking the Hermitian conjugate of $  \phi^{\dagger} i D_{\mu} \phi$ also adds a total divergence, so that we have 

\[   \mathcal{L}_{\text{Higgs}} = (( \hat{k}_{\phi} )^{\mu} + (\hat{k}_{\phi} )^{\dagger \mu}) \phi^{\dagger} i D_{\mu} \phi   + \frac{1}{2}  (  ( \hat{k}_{\phi D \phi} )^{\mu} + ( \hat{k}_{\phi D \phi} )^{\dagger \mu}  )  ( \phi^{\dagger} \phi ) ( \phi^{\dagger} i D_{\mu} \phi ) . \tag{4} \]

\noindent
The first term is minimal and was studied by Anderson, Sher and Turan, who stated that the one loop effects on the photon propagator due to this term can provide strong bounds on the coefficient $(\hat{k}_{\phi})^{\mu}$ [16].  We will provide a few more details in support of this statement.

 If we consider the covariant derivative $D_{\mu} \phi$ in the minimal term and suppress gauge indices, we have

\[ D_{\mu} \phi = \partial_{\mu} \phi - i [g_2 A^a_{\mu} T^a  + g_1 B_{\mu} Y]  \phi  ,\tag{5}\]

\noindent
where $Y$ is the hypercharge generator and $T^a = \frac{1}{2} \sigma^a$.  Note that the regular derivative $\partial_{\mu}$ should be replaced in this setting with the covariant derivative of general relativity (which we denote by $D^{*}_{\mu}$), but this is just a regular derivative because it acts on a scalar field.  In matrix form, equation (5) can be written as

\[ D_{\mu} \phi = \partial_{\mu} \phi - \frac{i}{2}  \begin{bmatrix}
g_2 A_{\mu}^3 - g_1 B_{\mu} & g_2 ( A_{\mu}^1 - i A_{\mu}^2 )\\
g_2 ( A_{\mu}^1 + i A_{\mu}^2 ) & - g_2 A_{\mu}^3 - g_1 B_{\mu}
\end{bmatrix}   \phi,  \tag{6} \]

\noindent
where the gauge indices have been suppressed again.  The physical fields are defined by

\[     W_{\mu}^{\pm} =  \frac{1}{\sqrt{2}} ( A_{\mu}^1   \mp i A_{\mu}^2  )       , \tag{7a}\]

\[ Z_{\mu} = c_w A_{\mu}^3 - s_w B_{\mu}                     ,  \tag{7b}\]

\[ A_{\mu} = s_w A_{\mu}^3 + c_w B_{\mu}                          ,\tag{7c}\]

\noindent
where $s_w$ and $c_w$ denote the sine and cosine of the Weinberg angle $\theta_w$, respectively.  The Higgs doublet $\phi$ is written in unitary gauge in the form $ \phi =  (v + \phi_1, 0 )^T / \sqrt{2}$, where $\phi_1$ is the Higgs field and $v$ is the corresponding vacuum expectation value.  Taking the interaction part of $((  \hat{k}_{\phi})^{\mu} + (\hat{k}_{\phi} )^{\dagger \mu} ) \phi^{\dagger} i D_{\mu} \phi $, we arrive at a term of the form

\[    \frac{1}{4}(( \hat{k}_{\phi} )^{\mu} + (\hat{k}_{\phi} )^{\dagger \mu}) \bigg[ ( g_2 s_w - g_1 c_w ) \phi_1 \phi_1 A_{\mu}   +   ( g_2 c_w + g_1 s_w ) \phi_1 \phi_1 Z_{\mu} \bigg] . \tag{8}     \]

\noindent
The first interaction vertex leads to the existence of a Higgs loop contribution to the vacuum polarization of the photon, where each vertex includes an insertion of a Lorentz-violating interaction.  By comparison with similar diagrams which appear for minimal terms with coefficients $\hat{k}_{\phi B}$ and $\hat{k}_{\phi W}$, it is expected that $\text{Re}((\hat{k}_{\phi})^{\mu}) \leq \mathcal{O} (10^{-16})$ GeV for all components.  On the other hand, modifications of electroweak symmetry breaking which occur with this term imply much stronger bounds ($\text{Re}((\hat{k}_{\phi})^{\mu}) \leq 10^{-31}$ GeV for the $X$ and $Y$ components and $\text{Re}((\hat{k}_{\phi})^{\mu}) \leq 10^{-27}$ GeV for the $Z$ and $T$ components) [16].  The symmetry breaking argument of Anderson et al. cannot be used for the term proportional to $( \hat{k}_{\phi D \phi} )^{ \mu}    ( \phi^{\dagger} \phi ) ( \phi^{\dagger} i D_{\mu} \phi ) $, but it is plausible that the coefficient should be of the same size or smaller than the coefficient for the minimal term, given that the analogue of equation (8) is identical apart from being higher order in the fields and having an additional two $\phi_1$ fields.  This would then give a bound $\text{Re}((\hat{k}_{\phi D \phi})^{\mu}) \leq 10^{-31}$ GeV$^{-1}$ for the $X$ and $Y$ components and $\text{Re}((\hat{k}_{\phi D \phi})^{\mu}) \leq 10^{-27}$ GeV$^{-1}$ for the $Z$ and $T$ components [17].

\subsection{Gauge Sector}
\noindent
We will next consider the gauge sector, which has four terms when we restrict to pseudovector backgrounds.

\subsubsection{Bounds on $\hat{k}_1$}

\noindent

\noindent
To begin, we consider the term $\big( \hat{k}_1  \big)_{\kappa}  \epsilon^{\kappa \lambda \mu \nu }  B_{\lambda} B_{\mu \nu} $.  Using 

\[ B_{\mu} = c_w A_{\mu} - s_w Z_{\mu}  , \tag{9a}\]

\[  B_{\mu \nu} = c_w F_{\mu \nu} - s_w Z_{\mu \nu}           ,\tag{9b}\]

\noindent
we find that one can have interactions of the form Z$\gamma \gamma$ , $\gamma$ZZ, $\gamma \gamma \gamma$, and ZZZ.  Triple interactions amongst the neutral gauge bosons are absent at tree level in the Standard Model and strongly suppressed at loop level, so interactions of this form are an important probe of new physics.  

To start with the Z$\gamma \gamma$ interaction, experimental searches for forbidden decays of the Z boson have established that the upper bound on the branching ratio for $\text{Z} \rightarrow \gamma \gamma$ is $1.46 \times 10^{-5}$ [18].  The term  we wrote above cannot be used to make predictions in a straightforward way because of the interdependence of the components of $\big( \hat{k}_1  \big)_{\kappa} $ on those of $B_{\lambda} B_{\mu \nu}$.  We may simplify by assuming that each component of the coupling constant vector $\big( \hat{k}_1  \big)_{\kappa} $ has the same value (denoted here by $K_1$) and the same bound.  The Lorentz-violating vertex function is then

\[  i \textbf{V}^{\nu \mu \rho}_{\text{Z} \gamma \gamma }  =     
- 16 i s_w c_w^2 \big[ (q-p)^{\rho} g^{\mu\nu} +  (r-q)^{\mu} g^{\nu \rho}  + (p -r)^{\nu} g^{\rho \mu} \big]   K_1           , \tag{10}  \]

\noindent
where $r=-p-q$ is defined to be the outgoing momentum of the second photon.

The corresponding decay amplitude is

\[   i \mathcal{T} =   - 16 i s_w c_w^2   \epsilon^{*}_{\nu} ( \lambda'_2, k'_2 )   \epsilon^{*}_{\rho} ( \lambda'_3, k'_3 )       \big[ (q-p)^{\rho} g^{\mu\nu}   +  (r-q)^{\mu} g^{\nu \rho} + (p -r)^{\nu} g^{\rho \mu} \big] \epsilon_{\mu} ( \lambda_1, k_1 )  K_1                ,\tag{11}\]

\noindent
where $\epsilon_{\mu}$ denotes a polarization of a particle.  Calculating the matrix element squared and summing over polarizations, we then have

\noindent

\[           \langle | \mathcal{T}|^2 \rangle =  1072 s_w^2 c_w^4 M_{\text{Z}}^2   K_1^2      .\tag{12}\]

\noindent
The decay width is  

\[  \Gamma_{\text{Z} \rightarrow \gamma \gamma}   = 21.32  s_w^2 c_w^4 M_{\text{Z}}  K_1^2       .\tag{13}\]

\noindent
Dividing through by the total decay width $\Gamma$ of the $\text{Z}$ boson, we have

\[ \text{Br} ( \text{Z} \rightarrow \gamma \gamma ) =  779.16   s_w^2 c_w^4   K_1^2   .\tag{14}\]

\noindent
Re-arranging for $K_1$, we find an upper bound on each component of $(\hat{k}_1)_{\mu} $ of $3.74 \times 10^{-4}$ GeV.  We note that this sensitivity is comparable to the one which was conjectured (but not proved) by Kostelecký  and Li for the coefficients $a^{(5) \mu \alpha \beta}$ in the limit of QCD and QED coupled to quarks, where it was suggested that the bound could be proved using direct simulations of the experimental effects on the cross section for deep inelastic scattering [6].

Moving on to the $\gamma \text{Z} \text{Z}$ interaction, there is no decay process in this case so we could instead consider the photon propagator.  Considering CPT-odd terms only, the photon Lagrangian can be written as

\[   \mathcal{L}_{\text{photon}} = - \frac{1}{4} F_{\mu \nu} F^{\mu \nu}  + \frac{1}{2} ( \hat{k}_{AF})^{\kappa} \epsilon_{\kappa \lambda \mu \nu} A^{\lambda} F^{\mu \nu}      . \tag{15} \]

\noindent
The equation of motion from this Lagrangian is 

\[  M^{\alpha \delta} A_{\delta} = 0 ,  \tag{16}    \]

\noindent
where

\[  M^{\alpha \delta} = g^{\alpha \delta} p^2 - p^{\alpha} p^{\delta} -2 i ( \hat{k}_{AF})_{\beta} \epsilon^{\alpha \beta \gamma \delta} p_{\gamma}    . \tag{17}\]

\noindent
To bound the coefficient, we can calculate the vacuum polarisation contributions to the photon propagator using the Lorentz-violating Lagrangian [16].  The result will be of the same form as the above equation and the value of the coefficient can be read off directly.  The $\hat{k}_{AF}$ may differ order by order in perturbation theory, so for simplicity we would have to restrict to divergent contributions to one-loop diagrams.  Cancellation of anomalies in the SME enforces the vanishing of $\hat{k}_{AF}$ at one loop, but this does not necessarily imply that $\hat{k}_1$ must also vanish at one loop, so we should instead consider contributions of this interaction to other processes.

One such process is ZZ production, which also includes a contribution from the anomalous ZZZ interaction.  Experimental measurements of these anomalous couplings are difficult because it would be involved in a process like $q \overline{q} \rightarrow$ ZZ which has many contributing diagrams in the Standard Model.  Measurements of such a coupling usually rely on an effective vertex approach, where the new couplings are assumed to conserve $U(1)$ and Lorentz invariance [19].  Calculation of the cross section with the Lorentz-violating term included would be complicated, since there is interference between the Lorentz-violating and conventional amplitudes, so we will leave this for future work in the interest of simplicity here.  It is not clear if this calculation would lead to a stronger bound than the one we derived using the decay rate, although this is possible.  One simplification would be to assume that the production cross section for a Lorentz-violating coupling is $q \overline{q}$ initiated only.  The Lorentz-violating vertex functions in this case would be

\[ i \textbf{V}^{\nu \mu \rho}_{\text{Z} \text{Z} \text{Z} }  =        
- 8 i s_w^3 \big[ (q-p)^{\rho} g^{\mu\nu} +  (r-q)^{\mu} g^{\nu \rho}  + (p -r)^{\nu} g^{\rho \mu} \big]  K_1     ,\tag{18a}\]

\[   i \textbf{V}^{\nu \mu \rho}_{\gamma \text{Z} \text{Z} }  =    4 i s_w^2 c_w \big[ (q-p)^{\rho} g^{\mu\nu} +  (r-q)^{\mu} g^{\nu \rho} + (p -r)^{\nu} g^{\rho \mu} \big]   K_1        .\tag{18b}\]

\noindent
Another simplification we will mention is that the quarks would be effectively massless, because we would be working in the high energy limit.  In this approximation, one could take the vector and axial vector coupling constants to be the average for the valence quarks which might be involved during a collision.  This would give us

\[ c_V - c_A \gamma^5 = - \frac{1}{3} \sin^2 \theta_w  . \tag{19} \]

\noindent
We may also do the same with the charges of the quarks which may be involved, which would give us

\[ Q_q = \frac{2}{3} e .\tag{20}\]

\subsubsection{ Bounds on $\hat{k}_2$}

\noindent
The arguments for the term with $\hat{k}_2$ are similar, but we will mention a few relevant points.  To begin, consider the parts of $\big( \hat{k}_2  \big)_{\kappa}  \epsilon^{\kappa \lambda \mu \nu }  \text{tr} ( W_{\lambda} W_{\mu \nu}   + \frac{2}{3} i g W_{\lambda} W_{\mu}  W_{\nu}) $ which involve $A_{\mu}^1$ and $A_{\mu}^2$.  Using equation (1) and

 \[  \frac{1}{\sqrt{2}} (F_{\mu \nu}^1 + i F_{\mu \nu}^2 ) = D_{\mu}^{\dagger} W_{\nu}^{-} - D_{\nu}^{\dagger} W_{\mu}^{-}           ,\tag{21a}\]

 \[  \frac{1}{\sqrt{2}} (F_{\mu \nu}^1 - i F_{\mu \nu}^2 ) = D_{\mu}W_{\nu}^{+} - D_{\nu} W_{\mu}^{+}           ,\tag{21b}\]

 \[ D_{\mu} W_{\nu}^{+} = D^{*}_{\mu} W_{\nu}^{+} - i g_2 (s_w A_{\mu} + c_w Z_{\mu}) W_{\nu}^{+}                   ,\tag{21c}\]

\noindent
we find that all the possible interactions involve an odd number of W bosons or some combination of a neutral gauge boson with an odd number of W bosons, so these terms are forbidden by conservation of charge.

If we instead consider the part which involves $A_{\mu}^3$ and use

\[  A_{\mu}^3 = s_w A_{\mu} + c_w Z_{\mu}        ,\tag{22a}\]

\[   F_{\mu \nu}^3 = s_w F_{\mu \nu} + c_w Z_{\mu \nu} - i g_2 ( W_{\mu}^{+} W_{\nu}^{-} - W_{\nu}^{+} W_{\mu}^{-}     )         ,\tag{22b}\]

\noindent
we find that there is again a Z $\rightarrow \gamma \gamma$ process, with the only difference being that the Lorentz-violating vertex function is now

 \[ i   \textbf{V}^{\nu \mu \rho}_{\text{Z} \gamma \gamma }  =        
 4 i s_w^2 c_w  \big[ (q-p)^{\rho} g^{\mu\nu} +  (r-q)^{\mu} g^{\nu \rho} + (p -r)^{\nu} g^{\rho \mu} \big]   K_2               , \tag{23}  \]

\noindent
with the same notation as before.  Following through with the calculation of the branching ratio, we reach a similar conclusion that $(\hat{k}_2)_{\mu} <  \mathcal{O} (10^{-4}$ GeV) for each component.

\subsubsection{Bounds on $\hat{k}_3$}

\noindent
There still remains the part involving the SU(3) gauge fields.  In this case,  the $(\hat{k}_3)_{\kappa} \epsilon^{\kappa \lambda \mu \nu} \text{tr} ( G_{\lambda} G_{\mu \nu})$ part yields a triple gluon vertex with a corresponding vertex function but a full comparison with experiment is complicated by the fact that one must consider all the possible interactions between quarks, antiquarks and gluons and combine the results with parton distribution functions to produce a prediction for the cross section for jet formation, which is beyond the scope of our work [20].  We note that there is only one diagram which contributes to this process at tree level if we use the Lorentz-violating vertex.  Gluon-gluon scattering would be more complicated and would also presumably not allow for a good bound, since two Lorentz-violating vertices would contribute four powers of the coupling constant $\hat{k}_3$ which would have to be cancelled out instead of two.  Although we are not able to bound these coefficients, we conjecture that they should have a sensitivity of order $10^{-4} $ GeV given currently available collider energies.  Our justification for this is that the same bound was found for the other coefficients $\hat{k}_1$ and $\hat{k}_2$ in the gauge sector and also that the same bound was suggested as plausible by Kostelecký and Li for the coefficients $a^{\mu \alpha \beta}$ in the limit of QCD and QED coupled to quarks [6].  

A dramatic qualitative prediction of Kostelecký and Li for the coefficients $a^{\mu \alpha \beta}$ is that there should be a difference between the differential cross section for deep inelastic scattering depending on whether a proton or an antiproton is used:

\[            \frac{d \sigma}{d x d y d \phi} \biggr\rvert_{e^{-}, p , a} \neq    \frac{d \sigma}{d x d y d \phi} \biggr\rvert_{e^{-}, \overline{p} , -a}  .  \tag{24}\]

\noindent
This follows because the coefficients govern CPT-odd operators, hence their contributions change sign when the proton is replaced with its antiparticle.   Since the same is true of $(\hat{k}_3)_{\kappa}$, one might ask if a similar effect is observable here.  The relevant leading order Feynman diagrams which are possible with the Lorentz-violating triple gluon vertex are shown in Figure 1.  Because of the CPT-odd operator, the contribution of the coefficient changes sign when a gluon is swapped with its antiparticle.  In high-energy scattering events, there may be processes where certain gluon exchanges dominate strongly, in which case the Lorentz-violating interaction suggests that the cross section should in principle be different depending on whether the gluon was exchanged or its antiparticle.  This may be interesting to explore in direct simulations, although as mentioned in the previous paragraph this would be beyond the scope of this paper.

\begin{figure}
  \centering
   \includegraphics[width=30mm]{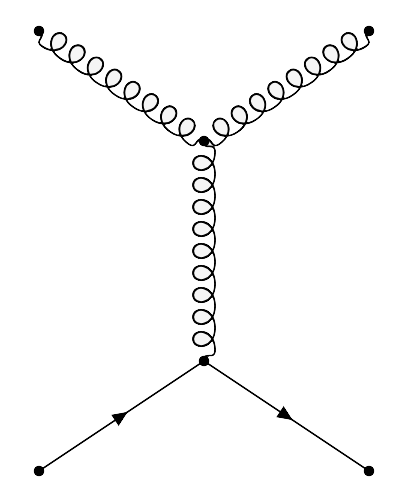}
    \includegraphics[width=50mm]{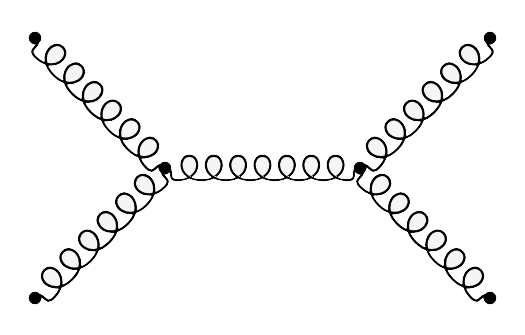}
  \includegraphics[width=40mm]{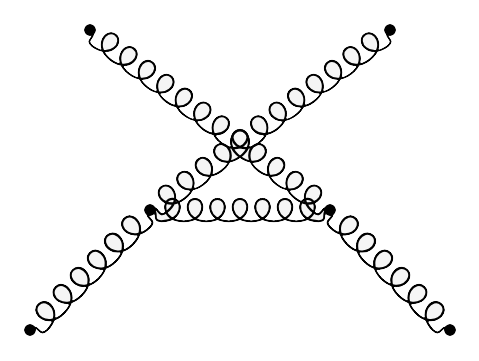}
 \includegraphics[width=30mm]{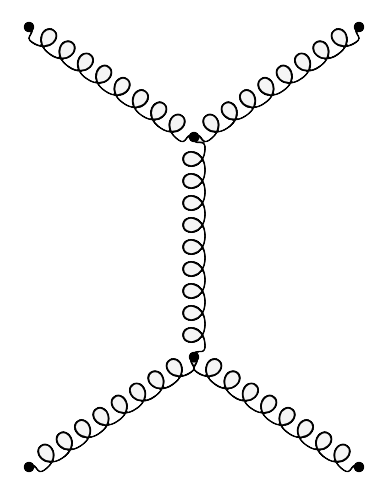}
  
       \caption{ Leading order diagrams for $q \overline{q} \rightarrow gg$ and $gg \rightarrow gg$ with a Lorentz-violating triple gluon vertex from the $(\hat{k}_3)_{\kappa} \epsilon^{\kappa \lambda \mu \nu} \text{tr} ( G_{\lambda} G_{\mu \nu})$  term.     }
     
\end{figure}

\subsubsection{Bounds on $\hat{k}_0$}

\noindent
We will finish by returning to the first term in $\mathcal{L}_{\text{gauge}}$, which is of the form $(\hat{k}_0 )_{\kappa} B_{\kappa}$.  One might naively expect this term not to contribute to any diagrams, since it only involves one field.  However, if we consider a scalar theory with an interaction term $j \phi$, where $j(x)$ is a classical field, this term can be treated as a perturbation which generates vertices in Feynman diagrams.  These linear terms are usually counterterms which are present to cancel tadpole graphs which appear at loop level, so we will bound $(\hat{k}_0)_{\mu}$ by considering it to be part of a term of this form.  The associated counterterm vertex factor for a Z boson is

\[  \textbf{V}^{\rho}  = - i s_w  (\hat{k}_0 )^{\rho}     . \tag{25}\]

\noindent
Using the Lorentz-violating ZZZ vertex, one can write an expression for a tadpole graph which contributes to a vacuum expectation value for $Z_{\mu}$.  This expression is set to be equal to the counterterm vertex factor:

\[  - i s_w  (\hat{k}_0 )^{\rho}   =  - 8 i s_w c_w^2 \big[ (q-p)^{\rho} g^{\mu\nu} +  (r-q)^{\mu} g^{\nu \rho} + (p -r)^{\nu} g^{\rho \mu} \big]  K_1 \int \frac{d^d k}{(2 \pi)^d}  \frac{  g_{\mu \nu} + \frac{k_{\mu} k_{\nu}}{M_Z^2} }{k^2 + M_Z^2 - i \epsilon}   .\tag{26}\]  

\noindent
The integral on the right-hand side is Lorentz-invariant, but the only Lorentz-invariant tensor of rank $2$ is the metric, so we must have

\[  - i s_w  (\hat{k}_0 )^{\rho}   =  -8i s_w c_w^2 \big[ (q-p)^{\rho} g^{\mu\nu} +  (r-q)^{\mu} g^{\nu \rho}   + (p -r)^{\nu} g^{\rho \mu} \big]  K_1 g_{\mu \nu}  \int \frac{d^d k }{(2 \pi)^d}  \frac{  1 + \frac{k^2}{d M_Z^2} }{k^2 + M_Z^2 - i \epsilon}   .\tag{27}\]  

\noindent
The right-hand side vanishes after tracing over the spacetime indices, which implies that all components of $(\hat{k}_0)_{\mu}$ vanish at one loop.

\subsection{Gravitational Sector}
\noindent

\noindent
The gravitational sector with single-index tensor coupling coefficients has three additional new terms, which we will repeat for convenience:

\[   \mathcal{L}_{\text{gravity}} =  \frac{1}{2 \kappa} \bigg[ ( \hat{k}_{\Gamma} )^{\mu}  \Gamma^{\alpha}_{\mu \alpha}  +    \big( \hat{k}_{\text{CS},1}  \big)_{\kappa}   \epsilon^{\kappa \lambda \mu \nu } \eta_{ac} \eta_{bd} \big( \omega_{\lambda}^{ab} \partial_{\mu} \omega_{\nu}^{cd}   + \frac{2}{3}  \omega_{\lambda}^{ab} \omega_{\mu}^{ce} \omega_{\nu e}^d  \big) \]  

\[+ \big( \hat{k}_{\text{CS},2}  \big)_{\kappa}   \epsilon^{\kappa \lambda \mu \nu }  \epsilon_{abcd} \big( \omega_{\lambda}^{ab} \partial_{\mu} \omega_{\nu}^{cd} + \frac{2}{3}  \omega_{\lambda}^{ab} \omega_{\mu}^{ce} \omega_{\nu e}^d  \big) \bigg]  .  \tag{28} \]

\noindent
We note that the first term is somewhat similar to the $\hat{k}_0$ term in the gauge sector if we were to view a set of Christoffel symbols as being analogous to a bare gauge field.  It is difficult to see how to bound these coefficients, since Lorentz violation in gravitational fields is usually considered only in the context of weak fields where the metric and vierbein are linearised [21].  Rather than working with approximately flat spacetimes, another possibility might be to work with a realistic spacetime which is simple to write down but which still has non-trivial curvature. An example is the Robertson-Walker metric, which is known to describe the current Universe up to a very good approximation:  

\[ ds^2 = dt^2 - a^2 \bigg( \frac{dr}{1 - k r^2} + r^2 ( d \theta^2 + \sin^2 \theta d \phi^2 ) \bigg) , \tag{29}\]

\noindent
where $k$ is the curvature parameter and $a$ is the scale factor.  Using this metric, the first term of $\mathcal{L}_{\text{gravity}}$ becomes

\[  \frac{1}{2 \kappa} ( \hat{k}_{\Gamma} )^{\mu}  \Gamma^{\alpha}_{\mu \alpha}  =  \frac{1}{2 \kappa} \bigg( ( \hat{k}_{\Gamma} )^{t}  \frac{\dot{a}}{a}  + ( \hat{k}_{\Gamma} )^{x} \bigg( \frac{2}{r} + \frac{kr}{1 - kr^2}  \bigg) +   ( \hat{k}_{\Gamma} )^{y} \cot \theta  \bigg) . \tag{30} \]

\noindent
 $\dot{a}/ a$ at our current time is equal to the Hubble constant $H_0$, which was recently measured to be $73.3^{+1.7}_{-1.8}$ km s$^{-1}$ Mpc$^{-1}$ using galactic lensing of quasars [22].   In units of s$^{-1}$, we may take $H_0 \approx 2.2 \times 10^{-18}$ s$^{-1}$.  The gravitational coupling constant is $1/ 2 \kappa \approx 3 \times 10^{36} $ GeV$^2$, so if we assume that the $t$ component of $(\hat{k}_{\Gamma})_{\mu}$ gives a term of order unity or lower when the coefficient multiplies the rest of the term, we find a bound on $(\hat{k}_{\Gamma})_{t}$ is of the order of $1 \times 10^{-18}$ GeV.     Repeating the above calculation with the Schwarzschild metric does not result in any useable bounds.  On the other hand, the coefficient of $(\hat{k}_{\Gamma})_{x}$ becomes infinite in the limit as $r$ goes to zero.  This could be due to the fact that the Robertson-Walker spacetime is not an exact description of the Universe or we could require that $(\hat{k}_{\Gamma})_{\mu}$ vanish to evade the infinity, since it is hard to see how it would be removed by using a more complicated cosmological metric.

We may instead consider the Chern-Simons-type terms.  A vierbein $e^{a}_{\mu}$ for the metric can be written down explicitly as

\[ e_0^0 =  1, \:\:\:\:\:\: e_1^1 =  \frac{a}{\sqrt{1 - k r^2}},   \:\:\:\:\:\: e_2^2 = ar, \:\:\:\:\:\: e_3^3 =  a r \sin \theta .\tag{31}\]

\noindent
The spin connection $\omega^{ab}_{\mu}$ is defined via

\[   \omega^{ab}_{\mu} = e_{\gamma}^a \Gamma^{\gamma}_{\sigma \mu} e^{\sigma b} + e_{\gamma}^a \partial_{\mu} e^{\gamma b}     .\tag{32}\]

\noindent
Using this to evaluate the Chern-Simons terms results in expressions which are complicated and not especially illuminating.  For example, we have 

\[ \frac{1}{2 \kappa} \big( \hat{k}_{\text{CS},1}  \big)_{0}   \epsilon^{0 \lambda \mu \nu } \eta_{ac} \eta_{bd} \omega_{\lambda}^{ab} \partial_{\mu} \omega_{\nu}^{cd}  = \frac{1}{2 \kappa} \big( \hat{k}_{\text{CS},1}  \big)_{t}  a r \bigg( 15 a^6 \dot{a} r^6 \cos \theta \sin^7 \theta   -15 a^6 \dot{a} r^5 \sin^8 \theta   \]

\[ + 3 a^7 r^8  (2kr^2 -3) \cos \theta \sin^7 \theta  -15 a^7 r^5 \cos \theta \sin^7 \theta + 6 a^7 r^5 \cos \theta \sin^6 \theta   -  \frac{3 a^7}{1 - kr^2} r^5 ( 5 kr^2 - 4) \cos \theta \sin^8 \theta \]

\[      - 12 a^3 r^2 \cos \theta \sin^2 \theta ( a^4 r^4 \sin^4 \theta \cos \theta - 3 a^4 r^3 \sin^3 \theta)                - 3 a^7 r^6 \sin^3 \theta ( 4 \cos^2 \theta \sin^3 \theta  - \sin^5 \theta) \]

\[+ 9 a^7 r^6 \cos^2 \theta \sin^6 \theta  + a^7 r^6 \cos^2 \theta  - 3 a^7 r^6 \sin^2 \theta \bigg). \tag{33}  \]

\noindent
The main point is that this term vanishes when $a$ or $r$ is zero, but not when $\dot{a}$, $k$ or $\theta$ are zero.  Since this is physically reasonable, we argue that it is plausible that $\big( \hat{k}_{\text{CS},1} \big)^{\mu}$ and $\big( \hat{k}_{\text{CS},2} \big)^{\mu}$ can both be taken as non-vanishing, with no a priori reason why we should set them to zero.

 \section{Spontaneous Diffeomorphism Violation and Nambu-Goldstone Modes  }

\noindent
An interesting possibility for Lorentz violation in the EFT is that it is determined by spontaneous breaking of Lorentz symmetry in the underlying quantum gravity theory.  This possibility is especially popular in string theory scenarios [23].  It is a general result that spontaneous violation of local Lorentz symmetry occurs if and only if there is also spontaneous violation of diffeomorphism invariance.    There is, however, an important exception.  This is when the tensor background fields are formed from combinations of the Minkowski metric $\eta_{ab}$ or the Levi-Civita tensor $\epsilon_{abcd}$, since in this case the fields can spontaneously break diffeomorphism invariance whilst accidentally preserving local Lorentz invariance.  The simplest example of terms of this kind can be found in the Higgs sector, since the tensor backgrounds $\hat{k}^{\mu \nu}$ could be proportional to the metric $\eta^{\mu \nu}$:

\[  \mathcal{L}_{\text{Higgs}} =  \frac{1}{2}  [   ( \hat{k}_{\phi \phi}  )^{\mu \nu}  (D_{\mu} \phi)^{\dagger} D_{\nu} \phi   + \text{H}. \text{c}. - ( \hat{k}_{\phi W} )^{\mu \nu}  \phi^{\dagger} W_{\mu \nu} \phi   - (\hat{k}_{\phi B})^{\mu \nu} B_{\mu \nu} \phi^{\dagger}  \phi      ]               , \tag{34}\]

\noindent
where the background fields have mass dimension $4$.  One can also have

\[ \mathcal{L}_{\text{Higgs}} = - ( \hat{k}_{\phi R})^{\mu \nu \rho \sigma}  R_{\mu \nu \rho \sigma} \phi^{\dagger} \phi  , \tag{35}\]

\noindent
since $(\hat{k}_{\phi R})^{\mu \nu \rho \sigma} $ could be proportional only to $\epsilon^{\mu \nu \rho \sigma}$ or $\eta^{\mu \nu} \eta^{\rho \sigma}$.

The existence of Nambu-Goldstone modes due to spontaneous Lorentz violation has been studied by Kostelecký and Bluhm [24].  In general, ten such modes can appear when a tensor acquires a vacuum expectation value, but only four of these are diffeomorphism modes associated with violation of diffeomorphism invariance.  The other six are Lorentz modes due to spontaneous breaking of local Lorentz invariance and would not be present for the backgrounds specified in the terms above.  However, the observability of these modes depends on the dynamics of the tensor field which causes the spontaneous breaking, as well as the geometry of the spacetime.  Given an arbitrary tensor $T_{ \lambda \mu \nu ...}$ and a vacuum value of the tensor $t_{\lambda \mu \nu ...}$, an excitation about the vacuum value is the difference between them:

\[ \delta T_{\lambda \mu \nu ..} = T_{ \lambda \mu \nu ...} - t_{\lambda \mu \nu ...}  . \tag{36}\]

\noindent
$T_{ \lambda \mu \nu ...}$ is some combination of $\eta_{\mu \nu}$ and $\epsilon_{\mu \nu \rho \sigma}$, so the vacuum value is a constant.  Since $T_{ \lambda \mu \nu ...}$ is also a constant, it follows that these fluctuations will not have any long-range effects and are not observable at long ranges.  Another way to see this is to consider the expression for the fluctuations due to infinitesimal diffeomorphisms

\[  \delta T_{\lambda \mu \nu ..} \approx - ( \partial_{\lambda} \zeta_{\alpha} ) t^{\alpha}_{\mu \nu} -   ( \partial_{\mu}  \zeta_{\alpha} ) t_{\lambda \nu}^{\alpha} - ... - \zeta^{\alpha} \partial_{\alpha} t_{\lambda \mu \nu}     ,  \tag{37} \]

 \noindent
 where $\zeta^{\mu}$ is a vector used to define an infinitesimal diffeomorphism in a coordinate basis:

 \[  x^{\mu}  \longrightarrow x^{\mu}  + \zeta^{\mu} . \tag{38}\]

 \noindent
More generally, in the case of constant vacuum values for tensors, one cannot form a kinetic term in the Lagrangian which allows propagation of diffeomorphism modes.  For this reason, in the rest of the article we do not distinguish between explicit and spontaneous diffeomorphism breaking, although the explicit case is usually ruled out by no-go theorems in Riemannian geometry [4, 25].

 \begin{table}[!h]
\begin{center}

\label{demo-table}
\begin{tabular}{||c c ||} 
 \hline
 Coefficient & Constraint   \\ [0.5ex] 
 \hline\hline
 $( \hat{k}_{\phi} )_{\mu}$ & $\text{Re}((\hat{k}_{\phi})_{X,Y}) \leq 10^{-31}$ GeV, $\text{Re}((\hat{k}_{\phi})_{Z,T}) \leq 10^{-27}$ GeV [8]   \\ 
 \hline
 $( \hat{k}_{\phi D \phi} )_{\mu}  $ & N/A but plausible $\text{Re}((\hat{k}_{\phi D \phi})_{X,Y}) \leq 10^{-31}$ GeV$^{-1}$, $\text{Re}((\hat{k}_{\phi D \phi})_{Z,T}) \leq 10^{-27}$ GeV$^{-1}$  \\
 \hline
 $(\hat{k}_0 )_{\mu}$ & $=0$ at one loop   \\
 \hline
 $\big( \hat{k}_1  \big)_{\mu} $ &  $<  10^{-4} \: \text{GeV}$  \\
 \hline
  $\big( \hat{k}_2  \big)_{\mu}$ & $< 10^{-4} \: \text{GeV}$   \\
 \hline
  $\big( \hat{k}_3  \big)_{\mu}$ & N/A but plausible $< 10^{-4}$ GeV  \\
 \hline
  $( \hat{k}_{\Gamma} )_{\mu} $ &  $( \hat{k}_{\Gamma} )_{t}< 1 \times 10^{-18}$ GeV   \\ [1ex] 
 \hline
  $\big( \hat{k}_{\text{CS},1}  \big)_{\mu}$ & N/A but plausible $\neq$ 0   \\ [1ex] 
 \hline
 $\big( \hat{k}_{\text{CS},2}  \big)_{\mu}$ & N/A but plausible $\neq$ 0   \\ [1ex] 
 \hline
\end{tabular}
\end{center}
\caption{Summary of results obtained for coefficients compared to previous work.}

\end{table}

  \section{Conclusion}

\noindent
To conclude, we will summarize the new results in the paper in systematic fashion and compare with previous studies.  The results obtained are shown in Table 1, where one can see the coefficient studied and the constraint obtained (where possible).  To begin with the Higgs sector, the coefficients $( \hat{k}_{\phi} )_{\mu}$ were already bounded by Anderson, Sher and Turan and we argued that it was plausible that the higher-order Higgs coefficients  $( \hat{k}_{\phi D \phi} )_{\mu}  $ should have the same bound [16].  In Section 2.B, we considered the four sets of coefficients which appear in the gauge sector, which have not been studied numerically before.  Bounds on $\big( \hat{k}_1  \big)_{\mu}$ and $\big( \hat{k}_2  \big)_{\mu}$ were obtained similarly to calculations of Kostelecký and Li by considering physical processes which would be predicted to be modified by these coefficients and then using experimental constraints on the process to bound the Lorentz-violating coefficients.  As an example, the presence of the term with $\big( \hat{k}_1  \big)_{\mu}$ predicts a modification to the branching ratio for the forbidden decay of a Z boson to a pair of photons which can be directly compared with experimental data.  We wish to emphasize the bounds which we found for $\big( \hat{k}_1  \big)_{\mu}$ and $\big( \hat{k}_2  \big)_{\mu}$ match the bound on the coefficients  $a^{\mu \alpha \beta}$ in the limit of QCD and QED coupled to quarks which was suggested as feasible by Kostelecký and Li given current experimental sensitivities [6].  The fact that we were able to consider different physical processes (for example, decay processes which are forbidden in the Standard Model) and use existing experimental measurements to match these bounds for the gauge sector suggests that this bound is likely to be correct for $a^{\mu \alpha \beta}$ given current data and simulations. 

Given the scope of this study, we were unable to bound the coefficients $\big( \hat{k}_3  \big)_{\mu}$ which take part in gluon interactions.  However, we noticed an interesting implication for processes dominated by certain gluon exchanges that the contribution from a coefficient changes sign when a gluon is exchanged with its antiparticle.  This is similar to the prediction of Kostelecký and Li that the presence of the $a^{\mu \alpha \beta}$ coefficients in QCD and QED coupled to quarks would ultimately lead to a difference between the deep inelastic scattering cross section for protons and antiprotons, although this would require further simulations to show conclusively.  Finally, we were also not able to bound the coefficients $\big( \hat{k}_0  \big)_{\mu}$ but by considering the predicted counterterm vertex factor for a Z boson, we were able to observe that the coefficients vanish at one loop (this has not been pointed out previously).

In Section 2.C, we studied the coefficients $( \hat{k}_{\Gamma} )_{\mu} $, $\big( \hat{k}_{\text{CS},1}  \big)_{\mu}$ and $\big( \hat{k}_{\text{CS},2} \big)_{\mu} $ in the gravitational sector.  These have not been bounded numerically in previous work, although Altschul showed that radiatively induced gravitational Chern-Simons terms due to a term in the fermion sector with a preferred axial vector are vanishing [26].  By assuming that the Universe is described to a very good approximation by the simple Robertson-Walker metric, the calculations for these terms are relatively tractable and result in a bound on the coefficient $( \hat{k}_{\Gamma} )_{t}$.  We also calculated an exact expression for one of the components of $\big( \hat{k}_{\text{CS},1}  \big)_{\mu}$ and argued that it was plausible that these coefficients do not vanish.

We note that the type of term studied in the article allows for multiple scenarios where diffeomorphism invariance may be explicitly or spontaneously broken without necessarily breaking local Lorentz invariance.  The case of explicit diffeomorphism violation is excluded by no-go theorems, but these theorems only apply to explicit breaking in Riemannian geometry, so there may still be loopholes if the breaking is due to a quantum gravity theory which is formulated in terms of non-Riemannian or Finsler geometry.  This could be another clue that the correct quantum gravity theory should be formulated using a type of geometry which is different from the familiar Riemannian one. It is also possible that the consistency conditions of the SME can be satisfied with explicit diffeomorphism breaking in a wide range of theories [27].  Explorations of gravity theories with new geometric settings are well-motivated even without quantum gravity, so this suggestion also fits in with wider work on modified gravity theories [28].

\section*{Acknowledgments}

\noindent
The author thanks Brett Altschul and Cosmas Zachos for helpful discussions.

\section*{References}

\noindent
[1] V. Alan Kostelecký  and Z. Li, Searches for beyond-Riemann gravity, Phys. Rev. D \textbf{104}, 044054 (2021).

\noindent
[2] S. Weinberg, Effective Field Theory, Past and Future, Proc. Sci. \textbf{CD09}, 001 (2009).

\noindent
[3] D. Colladay and V. A. Kostelecký, Lorentz-violating extension of the standard model, Phys. Rev. D \textbf{58}, 116002 (1998).

\noindent
[4] V. Alan Kostelecký, Gravity, Lorentz violation, and the standard model, Phys. Rev. D \textbf{69}, 105009 (2004).

\noindent
[5] B. Altschul, Gauge invariance and the Pauli-Villars regulator in Lorentz- and CPT-violating electrodynamics, Phys. Rev. D \textbf{70}, 101701 (2004).

\noindent
[6] V. Alan Kostelecký  and Z. Li, Gauge field theories with Lorentz-violating operators of arbitrary dimension, Phys. Rev. D \textbf{99}, 056016 (2019).
\newline
\noindent
V. Alan Kostelecký  and Z. Li, Backgrounds in gravitational effective field theory, Phys. Rev. D \textbf{103}, 024059 (2021).

\noindent
[7] V. Alan Kostelecký and N. Russell, Data tables for Lorentz and $CPT$ violation, Rev. Mod. Phys. \textbf{83}, 11 (2011).

V. Alan Kostelecký and N. Russell, Data tables for Lorentz and $CPT$ violation (update), arXiv:0801.0287v17.

\noindent
[8] O.W. Greenberg, $CPT$ Violation Implies Violation of Lorentz Invariance, Phys. Rev. Lett. \textbf{89}, 231602 (2002).

\noindent
[9] B.J. Powell, Emergent particles and gauge fields in quantum matter, Cont. Phys. \textbf{61}(2), 96-131 (2020).

\noindent
 [10] V. Alan Kostelecký et al., Lorentz violation in Dirac and Weyl semimetals, Phys. Rev. Res. \textbf{4}, 023106 (2022).

\noindent
[11] M.A. Ajaib, Lorentz violation and Condensed Matter Physics,  arXiv:1403.7622v2.

 \noindent
 [12] A. Gómez et al., Lorentz invariance violation and the CPT-odd electromagnetic response of a tilted anisotropic Weyl semimetal, Phys. Rev. D \textbf{109}, 065005 (2024).

\noindent
[13] A. Gómez, A. Martín-Ruiz and L.F. Urratia, Effective electromagnetic actions for Lorentz violating theories exhibiting the axial anomaly, Phys. Lett. B \textbf{829}, 137043 (2022).

\noindent
[14] R. D. Peccei and H. R. Quinn, CP conservation in the
presence of instantons, Phys. Rev. Lett. \textbf{38}, 1440 (1977).

\noindent
[15] B. Altschul, Bounds on vacuum-orthogonal Lorentz and CPT violation from radiative corrections, Phys. Rev. D \textbf{99}, 111701 (2019).

\noindent
[16] D.L. Anderson, M. Sher and I. Turan, Lorentz and $CPT$ violation in the Higgs sector, Phys. Rev. D \textbf{70}, 016001 (2004).

\noindent
[17] F. Canè et al., Bound on Lorentz and $CPT$ Violating Boost Effects for the Neutron, Phys. Rev. Lett. \textbf{93}, 230801 (2004).

\noindent
[18] T. Aaltonen et al. (CDF Collaboration), First Search for Exotic Z Boson Decays into Photons and Neutral Pions in Hadron Collisions, Phys. Rev. Lett. \textbf{112}, 111803 (2014).

\noindent
[19] S. Dutta, A. Goyal and Mamta, New physics contribution to neutral trilinear gauge boson couplings, Eur. Phys. J. C \textbf{63}, 305-315 (2009).

\noindent
[20] CFD Collaboration, Measurement of the diject mass distribution in $pp$ collisions at $\sqrt{s}= 1.8$ Tev, Phys. Rev. D \textbf{48}, 998 (1993).
\newline
\noindent
CMS Collaboration, Measurement of the triple-differential dijet cross section in proton-proton collisions at $\sqrt{s} = 8$ TeV and constraints on parton distribution functions, Eur. Phys. J. C \textbf{77}, 746 (2017).

\noindent
[21] Z. Li, Meausing Lorentz Violation in Weak Gravity Fields, arXiv:2301.12705v1. 

\noindent
[22] K.C. Wong,  A $2.4$ per cent measurement of $H_0$ from lensed quasars: 5.3$\sigma$ tension between early- and late-Universe probes, MNRAS \textbf{498}, 1420 (2020).

 \noindent
[23] V. Alan Kostelecký  and S. Samuel, Spontaneous breaking of Lorentz symmetry in string theory, Phys. Rev. D \textbf{39}, 683 (1989).

 \noindent
[24] R. Bluhm and V. Alan Kostelecký, Spontaneous Lorentz violation, Nambu-Goldstone modes, and gravity, Phys. Rev. D \textbf{71}, 065008 (2005).

\noindent
[25] J. Zhu and B.-Q. Ma, Lorentz Violation in Finsler Geometry, Symm. \textbf{2023}, 15(5), 978 (2023).
 
 \noindent
 [26] B. Altschul, There is no ambiguity in the radiatively induced gravitational Chern-Simons term, Phys. Rev. D \textbf{99}, 125009 (2019).

\noindent
[27] R. Bluhm and Y. Yang, Gravity with Explicit Diffeomorphism Breaking, Symm. \textbf{2021}, 13(4), 660 (2021).

 \noindent
[28] C.G. Boehmer and E. Jensko, Modified gravity: A unified approach, Phys. Rev. D \textbf{104}, 024010 (2021).

\end{document}